\documentclass[journal,12pt,onecolumn]{IEEEtran}

\ifCLASSINFOpdf
  \usepackage[pdftex]{graphicx}
  \graphicspath{{./figure/}}
  \DeclareGraphicsExtensions{.pdf,.jpg,.png}
\else
  \usepackage[dvips]{graphicx}
  \graphicspath{{./figure/}}
  \DeclareGraphicsExtensions{.eps}
\fi

\usepackage{algorithmic}
\usepackage{array}
\usepackage{amsmath,graphicx,booktabs}
\usepackage{multirow}
\usepackage{epstopdf,url}
\usepackage{makeidx,amssymb}  
\usepackage{caption}
\usepackage{comment}
\usepackage{color}
\ifCLASSOPTIONcompsoc
  \usepackage[caption=false,font=normalsize,labelfont=sf,textfont=sf]{subfig}
\else
  \usepackage[caption=false,font=footnotesize]{subfig}
\fi

\usepackage{url}
\usepackage{hyperref}       
\usepackage{amsmath,amsfonts,amssymb}
\usepackage{xcolor}

\newcommand{\revise}[1]{{\color{black}{#1}}}
\def\ie{\emph{i.e.}}
\def\eg{\emph{e.g.}}
\def\etal{{\em et al.}}

\begin{document}
\title{Metal Artifact Reduction in 2D CT Images with Self-supervised Cross-domain Learning}

\author{Lequan Yu,
	Zhicheng Zhang, 
	Xiaomeng Li,
	Hongyi Ren,
	Wei Zhao,
	and Lei Xing
	\thanks{L. Yu is with the Department of Statistics and Actuarial Science, The University of Hong Kong, Hong Kong, China, and also with the Department of Radiation Oncology, Stanford University, USA (email: lqyu@hku.hk).}
	\thanks{Z. Zhang, H. Ren, W. Zhao, and L. Xing are with the Department of Radiation Oncology, Stanford University, USA.}
	\thanks{X. Li is with Department of Electronic and Computer Engineering, Hong Kong University of Science and Technology, Hong Kong, China, and also with the Department of Radiation Oncology, Stanford University, USA.}
	\thanks{This work was partially supported by NIH (1 R01CA227713), Varian Medical Systems, and a Faculty Research Award from Google Inc. }
}

\maketitle

\begin{abstract}
The presence of metallic implants often introduces severe metal artifacts in the X-ray CT images, which could adversely influence clinical diagnosis or dose calculation in radiation therapy. In this work, we present a novel deep-learning-based approach for metal artifact reduction (MAR). In order to alleviate the need for anatomically identical CT image pairs (\ie, metal artifact-corrupted CT image and metal artifact-free CT image) for network learning, we propose a self-supervised cross-domain learning framework. Specifically, we  train a neural network to restore the metal trace region values in the given metal-free sinogram, where the metal trace is identified by the forward projection of metal masks. We then design a novel FBP \revise{reconstruction} loss to encourage the network to generate \revise{more perfect} completion results and a residual-learning-based image refinement module to reduce the secondary artifacts in the reconstructed CT images. To preserve the fine structure details and fidelity of the final MAR image, instead of directly adopting CNN-refined images as output, we incorporate the metal trace replacement into our framework and replace the metal-affected projections of the original sinogram with the prior sinogram generated by the forward projection of the CNN output. We then use the filtered backward projection (FBP) algorithms for final MAR image reconstruction. We conduct \revise{an} extensive evaluation on simulated and real artifact data to show the effectiveness of our design. Our method produces superior MAR results and outperforms other compelling methods.
We also demonstrate the potential of our framework for other organ sites.

\end{abstract}
\section{Introduction}
In modern healthcare \revise{procedures}, computed tomography (CT) images are widely used for medical diagnosis and treatment. 
However, in clinical practice, patients often carry highly attenuated metallic implants, \eg, hip prostheses, and such metals will lead to unreliable X-ray projections.
If we directly reconstruct CT images from those unreliable X-ray projections, the reconstructed CT images will present strong metal artifacts, which will \revise{affect} not only the visual diagnosis from CT images but also the dose calculation in radiation therapy~\cite{kalender1987reduction,meng2010sinogram}.
In the past decades, plenty of metal artifact reduction (MAR) methods have been developed to improve the quality of CT images. 
However, there is a large gap between in-house research and real clinical practice~\cite{gjesteby2016metal}. How to conduct metal artifact reduction is still an important yet \revise{challenging} problem in the \revise{research field}~\cite{park2018ct}.

Considering the metal artifacts are structured and non-local in  CT images, the previous methods conduct MAR from the raw X-ray projection level (\ie, sinogram).
In these works, the metal-affected projections are either corrected by modeling the physical effects of CT imaging~\cite{hsieh2000iterative,kachelriess2001generalized,meyer2010empirical,park2015metal}, or replaced by estimated values~\cite{kalender1987reduction,meyer2010normalized,zhang2011efficient,zhang2011new,zhang2013hybrid,mehranian2013x}.
The linear interpolation (LI)~\cite{kalender1987reduction} is one typical solution, which regards the metal-affected sinogram regions as missing and directly \revise{replaces} the missing projection by linear interpolation of its neighboring unaffected projections. 
However, the inconsistency between interpolated and unaffected values often results in strong secondary artifacts in the reconstructed images.
To alleviate this problem, some prior-image-based interpolation methods~~\cite{bal2006metal,muller2009spurious,prell2009novel, meyer2010normalized, wang2013metal,zhang2013hybrid,zhang2014metal} were proposed.
The basic idea is to reconstruct a good prior image (sinogram) from the metal artifact images and then use the prior sinogram to guide the sinogram interpolation.
For example, \cite{meyer2010normalized} proposed to use a multithreshold segmentation method to acquire the prior image and then normalize the projection data before interpolation.

Inspired by the success of convolutional neural networks (CNNs) in medical image reconstruction and analysis~\cite{jin2017deep, chen2017low,chen2017low2, wolterink2017generative,wang2018image,zhang2018sparse,litjens2017survey}, \revise{recent works have been proposed to address MAR with CNNs~\cite{xu2018deep,ghani2019fast,lin2019dudonet,liao2019generative,liao2019adn,lyu2020dudonet,wang2021dan}.}
Park~\etal~\cite{park2018ct} applied U-Net~\cite{ronneberger2015u} to correct sinogram inconsistency, while Gjesteby~\etal~\cite{gjesteby2017deep} used neural networks to improve the NMAR~\cite{meyer2010normalized} approach in the sinogram domain.
Besides addressing MAR on the sinogram domain with deep learning, some other works reduce the metal artifacts via image post-processing~\cite{park2017machine,gjesteby2017reducing,gjesteby2017deep,gjesteby2018deep}.
Huang~\etal~\cite{huang2018metal} employed a novel residual learning method based on CNN to reduce metal artifacts in cervical CT images\revise{, and} Gjesteby~\cite{gjesteby2018deep} additionally adopted a perceptual loss to improve the quality of residual-learning-based MAR results. 
There are also some works adopting the conditional generative adversarial network (cGAN)~\cite{isola2017image} for MAR~\cite{wang2018conditional}.
Meanwhile, Zhang~\etal~\cite{zhang2018convolutional} employed CNN to generate a high-quality prior image from raw and other methods corrected CT images and then \revise{used} this prior image to help correct the metal artifacts, while our recent work~\cite{yu2020deep} further improved this framework by using CNN to conduct prior-image-based sinogram completion.

Although the above deep-learning-based MAR approaches achieved good performance in their experimental data, most of them are supervised approaches and require anatomically identical CT image pairs for network training, which consist of metal artifact corrupted CT image and metal artifact free CT image.
Considering it is difficult or impractical to collect such training pairs, most of these methods simulate metal-affected CT images from metal-free CT images to obtain training pairs.
However, the synthesized metal artifacts are difficult to reproduce the clinical scenarios and the performance of these supervised methods would degrade in real clinical practice~\cite{liao2019adn}.
Therefore, very recent works focus on develop MAR methods without utilizing the synthesized metal artifacts.
Liao~\etal~\cite{liao2019adn} proposed an unsupervised artifact disentanglement network to disentangle the metal artifacts from CT images in the latent space.
Ghani~\etal~\cite{ghani2019fast} treated the projection data corresponding to metal objects as missing data and proposed a self-supervised sinogram inpainting framework to complete the missing data in the projection domain.
However, this method only focuses on the sinogram completion results without considering the secondary artifacts in the reconstructed images.

\begin{figure*}[t]
	\centering
	\includegraphics[width=1.0\linewidth]{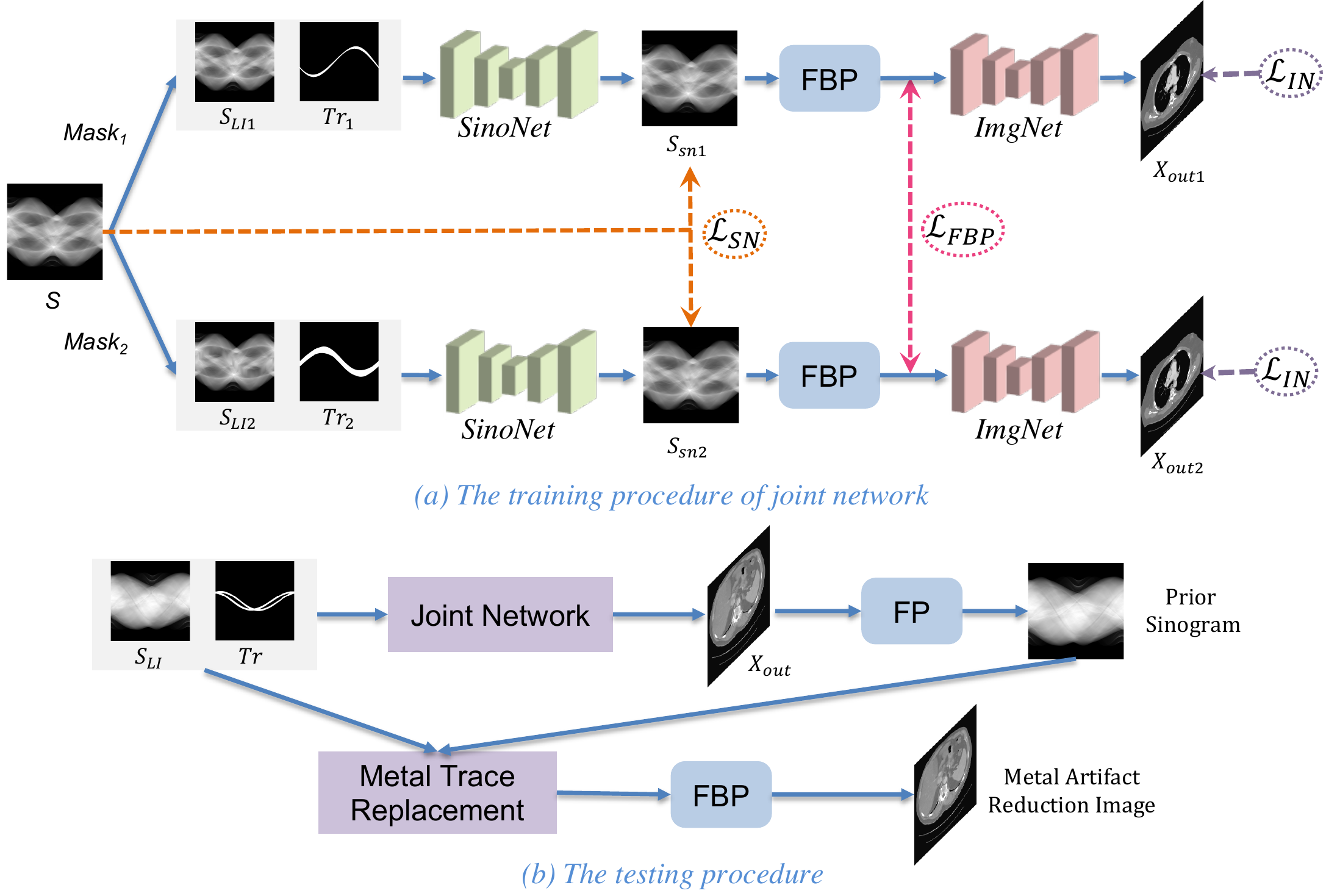}
	\caption{Illustration of the proposed self-supervised cross-domain learning framework. (a) We train a sinogram completion network, \ie, SinoNet and an image refinement network, \ie, ImageNet in an unsupervised manner. 
	In the testing phase (b), we use the forward projection (FP) of CNN output as prior sinogram to replace the metal-affected values in \revise{the} original sinogram, and use the FBP algorithm to reconstruct the final metal artifact reduced CT image from the replaced sinogram.}
	\label{fig:network_train}
	\centering
\end{figure*}

In this work, we present a novel self-supervised deep-learning-based framework for metal artifact reduction to alleviate the need for anatomically identical metal-free and metal-corrupted CT image pairs.
In particular, given a raw metal-free sinogram, we simulate the metal trace with a forward projection of randomly selected metal masks, and then train a network (\ie, SinoNet) to restore the projection values in the metal trace region.
Different from the previous sinogram inpainting method, we design a novel FBP \revise{reconstruction} loss to encourage the network to generate \revise{more perfect sinogram} completion results by minimizing the difference of a pair of completion results from the same sinogram.
Since direct sinogram completion would often lead to secondary artifacts in the reconstructed CT images, we then apply a residual-learning-based image refinement module to reduce the secondary artifacts in the image domain.
Although the metal artifacts are significantly reduced with the above procedures, some tiny anatomical structures in the CNN-refined images would change. 
Therefore, to preserve the fine-grained details of final MAR images, we incorporate metal trace replacement into our framework and replace the metal-affected projections of the original sinogram with the forward projection of CNN output. 
After that, the final MAR images are directly FBP-reconstructed from the replaced sinogram to preserve the fidelity and low-contrast features.
Notably, our method only needs metal-free sinogram and metal masks for network training, and thus alleviate the need for synthesizing metal artifacts.

Our main contributions are summarized as follows.
\begin{itemize}
	\item We propose a new self-supervised learning framework to conduct MAR without requiring metal-free and metal-corrupted training pairs via cross-domain sinogram and image learning.

	\item We design a novel FBP \revise{reconstruction} loss to encourage the network to generate \revise{more perfect sinogram} completion results.
	We also incorporate the reconstruction model (metal trace replacement) into our framework as the last step to preserve fine details and avoid resolution loss.
	
	\item We evaluate our method on CT images with simulated metal artifacts and real metal artifacts to show the superior MAR results. The experiment on other organ images shows the potential of our method to generalize to other sites.
\end{itemize}

The rest of this article is described as follows. Section~\ref{sec:method} illustrates the details of our self-supervised framework for MARI. Section~\ref{sec:experiment} shows the experimental setting and results. Section~\ref{sec:discussion} and Section~\ref{sec:conclusion} presents the discussion and conclusion of this study, respectively.
\section{Method}
\label{sec:method}

Fig.~\ref{fig:network_train} illustrates the overview of our  proposed self-supervised learning framework for metal artifact reduction in CT images.
We develop a self-supervised learning framework to train a SinoNet to complete the missing region (\ie, metal trace) of the sinogram.
An FBP consistency loss is adopted to guide the SinoNet to generate geometry-consistent results by encouraging the consistency of FBP reconstructions from a pair of completion results.
Also, an image refinement module, \ie, ImgNet, is employed to reduce the secondary artifacts in reconstructed images.
In the testing phase, instead of directly taking the CNN-refined images as output, we use metal trace replacement to generate the final completed sinogram  and the MAR image is reconstructed from the completed sinogram, so that we can preserve the fine structure details and fidelity of the final MAR image

\subsection{Self-supervised Cross-domain Learning}

\subsubsection{Sinogram completion}
In this step, we train a network (referred to \textit{SinoNet}) to complete the missing projection data in \revise{the} metal trace.
The SinoNet takes the metal-corrupted  sinogram $S_{ma} \in \mathbb{R}^{N\times D}$ and the  corresponding metal trace  $Tr \in \{0,1\}^{N\times D}$ as inputs, and \revise{estimates} the projection values in \revise{the} metal trace region of $S_{ma}$, where $N$ is the number of projection views and $D$ is the number of detectors.
Considering that it is difficult to acquire the pairs of metal-corrupted sinogram and metal-free sinogram to optimize the network, we propose a self-supervised learning manner to learn the sinogram completion network.
Specifically, we only employ the metal-free sinogram data to train the network.
To do that, given a metal-free sinogram $S$, we first simulate a metal trace  $Tr$ from a pre-collected metal mask $M$ as
\begin{equation}
Tr = \mathcal{P}(M) >0, 
\end{equation}
where $\mathcal{P}$  is the forward projection.
We then replace the values within the metal trace as zero in $S$ to get the incomplete sinogram $S_{ma}$  to simulate the metal-corrupted sinogram.
Then we train the network to restore $S$ from $S_{ma}$.

To ease the network learning, we employ linear interpolation to produce an initial estimation for the missing sinogram data and take the interpolated sinogram $S_{LI}$ and $Tr$ as network input.
Considering the metals only cause unreliable projection data in \revise{the} metal trace region, we then composite the results of SinoNet and interpolated sinogram $S_{LI}$ with respect to $Tr$ to acquire the completion result:
\begin{equation}
S_{sn} = f_{S}(S_{LI}, Tr)\odot Tr + S_{LI} \odot (1-Tr),
\end{equation}
where $f_{S}$ represents the procedure of SinoNet and $\odot$ is element-wise multiplication. 
To encourage the network to extract more representative features at \revise{the} metal trace region for network learning, we adopt the mask pyramid U-Net~\cite{liao2019generative} as the network backbone. 
This network architecture explicitly incorporates the metal information into each layer and thereby \revise{guides} the sinogram completion network to learn discriminative features for sinogram completion.

\subsubsection{\revise{FBP reconstruction loss}}
To optimize the SinoNet, we adopt L1 loss to minimize the difference between completion result $S_{sn}$ and the metal-free sinogram $S$:
\begin{equation}
\mathcal{L}_{SN} = || S_{sn} - S||_1.
\end{equation}
However, the L1 loss only penalizes the inconsistency of individual projection values within the metal trace, \revise{without considering the potential secondary artifacts caused by imperfect sinogram completion}.

Therefore, we  design a new \revise{FBP reconstruction loss}, encouraging the SinoNet to generate \revise{more perfect} completion to alleviate the secondary artifacts.

Given one metal-free sinogram $S$, we generate a pair of metal-corrupted sinogram $S_{ma1}$ and $S_{ma2}$ with respect to different metal \revise{traces} $Tr_1$ and $Tr_2$.
We feed the pair of metal-corrupted sinograms into SinoNet and get two completed sinogram $S_{sn1}$ and $S_{sn2}$.
As these two completed sinograms are from the same metal-free sinogram $S$, the corresponding FBP reconstructions should be the same. 
Therefore, we minimize the L1 difference between two FBP reconstructed results:
\begin{equation}
\begin{aligned}
\mathcal{L}_{FBP} &=||(\mathcal{P}^{-1}(S_{sn1})- \mathcal{P}^{-1}(S_{sn2}))\odot (1 - M_1 | M_2)||_1,
\end{aligned}
\end{equation}
where $\mathcal{P}^{-1}$ denotes the FBP operation,  $M_1$ and $M_2$ are the metal masks.
Since it is difficult to reconstruct accurate CT images in metal regions,  we employ a masked L1 loss to minimize only the non-metal regions of CT images.
Note that we implement the FBP operation $\mathcal{P}^{-1}$ in a \textit{differentiable} manner, so that the gradient of $\mathcal{L}_{FBP}$ can back-propagate to SinoNet to enable network training.

\subsubsection{Image refinement}
%
To further reduce the secondary artifacts in the reconstructed images, we apply an image refinement module (refereed to \textit{ImgNet}) to refine the reconstructed CT image.
Given the completed sinogram $S_{sn}$, the refined image can be represented as:
\begin{equation}
X_{out} = \mathcal{P}^{-1}(S_{sn}) + f_{I} (\mathcal{P}^{-1}(S_{sn})),
\end{equation}
where $f_{I}$ denotes the image refinement network.
The image refinement module follows the U-Net architecture with halving the channel number of each layer.
We need the ground truth \revise{reference images} to optimize the network, while in our framework, we only have \revise{a} metal-free sinogram.
Therefore, we first employ the FBP operation on the metal-free sinogram $S$ to acquire an FBP reconstruction as the ``\revise{reference} image" and then use a LI loss between the network output and the FBP reconstruction
\begin{equation}
\mathcal{L}_{IN} = || X_{out} - \mathcal{P}^{-1}(S)||_1.
\end{equation}

We jointly train the \textit{SinoNet} and \textit{ImgNet}, and the full objective function is
\begin{equation}
\mathcal{L} _{total}= \mathcal{L} _{SN} + \alpha_{1}\mathcal{L} _{FBP}+ \alpha_{2}\mathcal{L} _{IN},
\end{equation}
where $\alpha_{1}$ and $\alpha_{2}$ are hyper-parameters and we empirically set them as 1.0.

\subsection{Prior-image-based Metal Trace Replacement}
With the above procedures, the metal artifacts are significantly reduced. 
However, in our preliminary experiments~\cite{yu2020deep}, we found that the CNN-refined images are often blurred and some tiny structures are lost. 
Inspired by the prior-image-based MRA methods, we incorporate a metal trace replacement into our framework to generate the final MAR image.
As shown in Fig.~\ref{fig:network_train} (b), we utilize the forward projection $S_{prior}$ of CNN-refined image $X_{out}$ to replace the metal-affected projections in \revise{the} original sinogram to get the corrected sinogram $S_{corr}$, and the final metal-free image is directly reconstructed from  $S_{corr}$.
Specifically, to eliminate the discontinuity at the boundary of the metal trace when conducting metal trace replacement, We conducted linear interpolation on the residual sinogram $S_{res}$ and the final corrected sinogram $S_{corr}$ can be represented as:
\begin{align}
\centering
S_{res} &= LI(S_{LI} -S_{prior}, Tr),  \nonumber \\
S_{corr} &= S_{prior} + S_{res}, 
\end{align}
where $LI$ denotes the linear interpolation.
Specifically, the $LI$ operation takes a sinogram $S\in \mathbb{R}^{N\times D}$ and a metal trace $Tr\in\{0,1\}^{N\times D}$ as input. 
The  $k$th projection pixel of $i$th view in the sinogram data is denoted as $S_{i,k}$. The $LI$ interpolation is performed in each projection view $S_i$. For $S_i$, if  $\{S_{i,k}| k \in [j+1,j+\Delta]\}$ are affected by metal objects (\ie,$Tr_{i,k}=1$) and its neighboring projection pixels $S_{i,j}, S_{i,j+\Delta+1}$ are not affected by metal objects, \revise{the projection pixels} $S_{i,k}$ are obtained by
\begin{equation}
S_{i,k} = S_{i,j}+\frac{S_{i,j+\Delta+1}-S_{i,j}}{\Delta+1}(k-j).
\end{equation}
Note that the metal-unaffected projections in $S_{corr}$ have the same value as in $S_{ma}$ (or $S_{LI}$) and the replaced data in the metal trace can seamlessly connect to the unaffected neighborhoods.
Finally, we use the conventional FBP algorithm to reconstruct the MAR image from the corrected sinogram, preserving the fidelity of the final image.

\subsection{Training and Testing Strategies}
To train our model, we only need raw metal-free sinograms and metal masks to generate  metal traces, alleviating the need  for simulated CT image pairs with and without metal artifacts.
In the inference phase, we first use a threshold-based method to segment the metal mask $M$  from the FBP-reconstructions of the metal-affected sinogram  $S_{ma}$. We then forward project the segment metals to acquire metal trace $Tr$.
We view the projection data with the metal trace as missing and acquire $S_{LI}$ with the linear interpolation.
After that, we can get the final MAR image with our proposed method.

\section{Experiments}
\label{sec:experiment}

\subsection{Datasets}
Due to \revise{the} lack of relevant clinical databases,  we trained our network with the simulated sinogram from DeepLesion~\cite{yan2018deep} CT images. 
To simulate the possible metal traces, we used the masks in~\cite{zhang2018convolutional}, which contains 100 masks with different metal shapes and sizes, such as hip prostheses, spine fixation screws, coiling, wires, etc.
For network training, we randomly chose 1000 CT slices from the DeepLesion dataset and 90 metal masks from the collected metal masks. Another 10 metal masks were combined with extra 200 CT slices to generate 2000 combinations for evaluation.
To simulate the metal-free sinogram, we followed the setting in~\cite{zhang2018convolutional} and used a fan-beam geometry with a polychromatic X-ray source.
During the simulation, all images and metal masks are resized to $416\times416$. The sizes of the metal objects in training data range from 16 to 4967 pixels, while the sizes of metal objects in testing data range from 32 to 2054 pixels.
We used 640 projection views and the sinogram size is $640\times641$. 
We follow the procedure in~\cite{zhang2018convolutional} to synthesize metal-affected sinograms by inserting metallic implants into clean CT images.
Note that the synthesized metal-affected CT images are only used for comparison with other supervised learning methods and we do not employ such training pairs to train our framework.

\subsection{Implementation Details}
We implemented our framework in Python with PyTorch~\cite{paszke2019pytorch} library.
The SinoNet and ImgNet were trained in an end-to-end manner with a differential FBP operation in \revise{the} ODL library\footnote{\url{https://github.com/odlgroup/odl}}. 
In total, we trained 180 epochs with a mini-batch size of 4 on one Nvidia 1080Ti GPU.
We used the Adam optimizer~\cite{kingma2014adam} with $(\beta1, \beta2)=(0.5, 0.999)$ to optimize the whole framework. 
We set the learning rate as $1e^{-4}$. 
In each training iteration, we sampled one mini-batch CT images and randomly paired with two different metal mask mini-batches to feed into the network.
We employ the root mean square error (RMSE) and structured similarity index (SSIM) metrics to quantitatively evaluate the performance of our method. 

\begin{table} [t]
	\centering
	\caption{Ablation study on different components of our framework.}
	\label{Table:ablation}
	
	\begin{tabular}{p{3.7cm}<{\centering}|p{1.9cm}<{\centering}|p{1.9cm}<{\centering}}
		\toprule[1pt]
		Method & RMSE (HU) & SSIM\\ \hline	
		SinoNet				     	&42.60     &0.9727     		\\
		SinoNet+FBPcons   	&41.97     &0.9731    	\\
		JointLearning	      	&36.28     &0.9755    	  \\
		JointLearning+MTR \textbf{(Ours)}            &\textbf{34.20}     &\textbf{0.9773}    \\	
		\bottomrule[1pt]
	\end{tabular}
\end{table}
\begin{figure*}[t]
	\centering
	\includegraphics[width=1.0\linewidth]{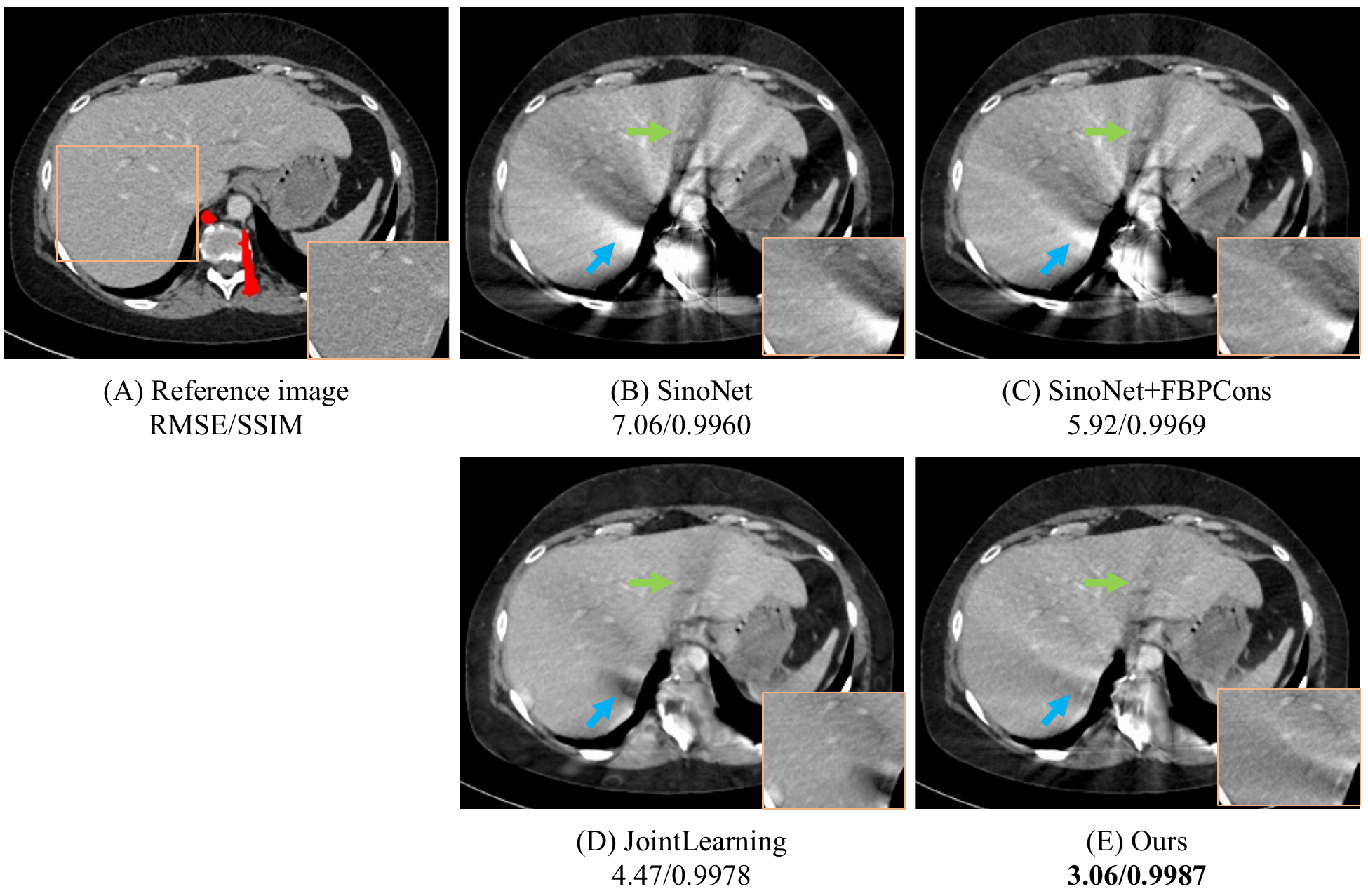}
	\caption{The visual comparison of different methods for ablation study. The metal masks are remarked as red in the \revise{reference} image. The numbers are ROI metrics of the orange patches for quantitative comparison. The display window is [-140 240] HU.
	\label{fig:ablation}}
	\centering
\end{figure*}
\subsection{Ablation Study}
We conducted a set of ablation study experiments to evaluate the effectiveness of each component: (1) only sinogram completion results without FBP consistency loss (SinoNet), (ii) sinogram completion results with proposed FBP \revise{reconstruction} loss (SinoNet+ FBPcons), (iii) the joint learning results of sinogram completion and image refinement (JointLearning), and (iv) our proposed method with metal trace replacement (JointLearning+MTR). For the method (i) and (ii), the final CT image is FBP-reconstructed from the completed sinogram. 
\begin{table} [t]
	\centering
	\caption{Comparison of different methods on DeepLesion simulation dataset.}
	\label{table:deeplesion}
	\begin{tabular}{p{3cm}<{\centering}|p{2cm}<{\centering}|p{2cm}<{\centering}}
		\toprule[1pt]
		Method & RMSE (HU) & SSIM\\ \hline
		
		LI~\cite{kalender1987reduction}      		 &50.31      &0.9455     \\
		NMAR~\cite{meyer2010normalized} 	    &47.03     &0.9594	   \\
		CNNMAR~\cite{zhang2018convolutional}&43.27     &0.9706 	   \\
		cGANMAR~\cite{wang2018conditional}         & 39.01	&0.9754  \\
		DuDoNet~\cite{lin2019dudonet}     &38.00      &0.9766 \\ 
		\textbf{Ours }      &\textbf{34.20 }    &\textbf{0.9773}			\\		
		\bottomrule[1pt]
	\end{tabular}
\end{table}
\begin{figure*}[h]
	\centering
	\includegraphics[width=1.0\linewidth]{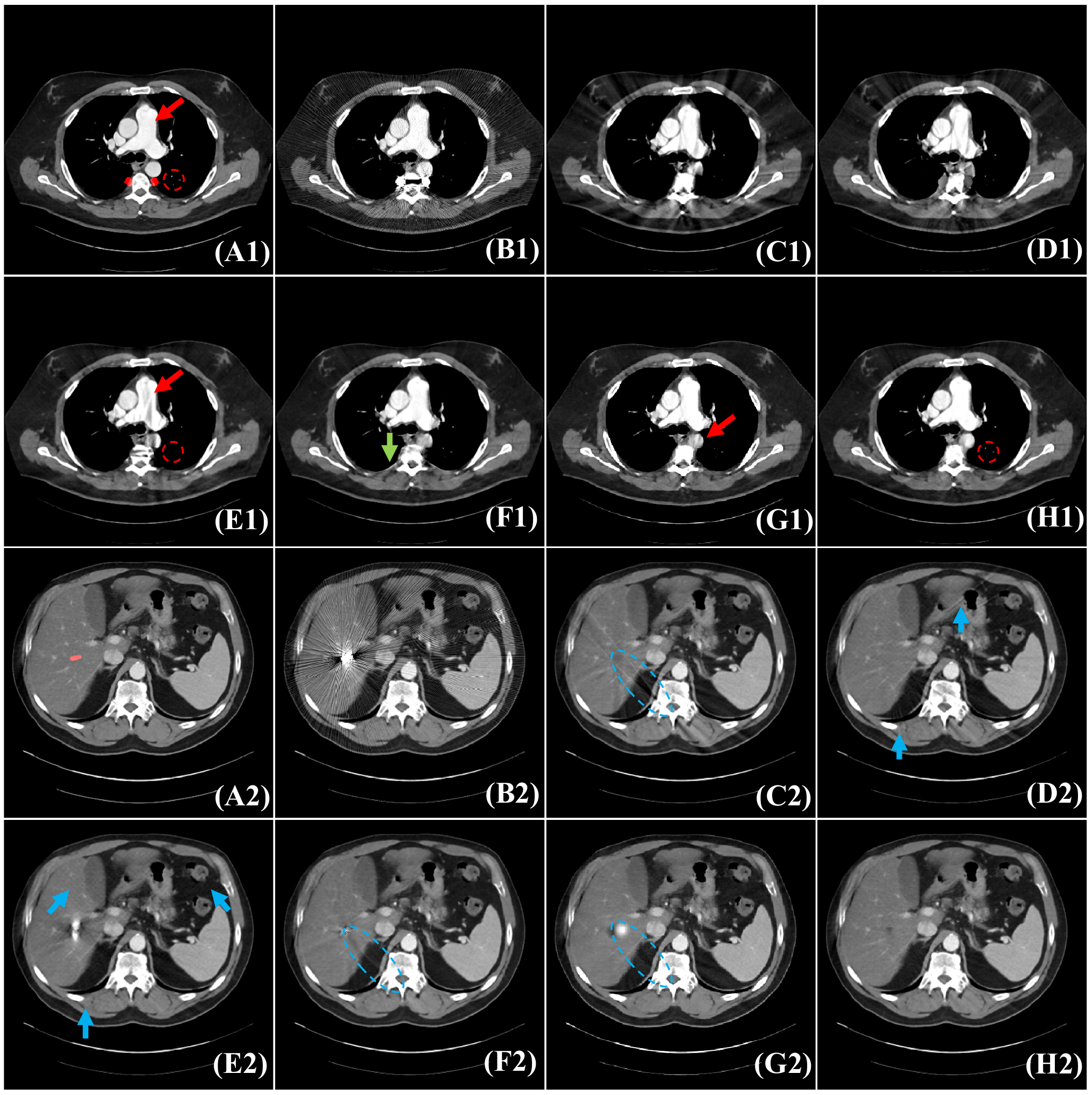}
	\caption{The qualitative results on DeepLesion  images from chest (A1-H1) and abdomen (A2-H2). We show \revise{reference images} (A1, A2) and the corresponding metal-corrupted CT images (B1, B2). We compare our method (H1, H2) with LI (C1, C2), NMAR (D1, D2), CNNMAR (E1, E2), cGANMAR(F1, F2), and DuDoNet (G1, G2). The figures are shown in [-200 200] HU.}
	\label{fig:visual1}
	\centering
\end{figure*}
Note that the results of SinoNet can be regarded as the sinogram inpainting method.
The results of different ablation models are shown in Table~\ref{Table:ablation}.
With the proposed FBP \revise{reconstruction} loss, the sinogram completion method improves performance on the DeepLesion simulation dataset compared with SinoNet. 
By incorporating the image domain refinement, the joint learning  outperforms the sinogram completion methods, showing the effectiveness of joint two domain learning. 
Compared with JointLearning, our method further reduces 3.74 HU  on the RMSE metric, confirming the effectiveness of metal trace replacement on preserving the fidelity of the final MAR image.
We also present an visual comparison for different methods for ablation study.
As shown in Fig.~\ref{fig:ablation}, the network with FBP \revise{reconstruction} loss reduces the severe artifacts near the spine regions compared with SinoNet; see the blue arrows in Fig.~\ref{fig:ablation}(B\&C).  Our method further reduces the artifacts compared with JointLearning (see the green and blue arrows in Fig.~\ref{fig:ablation}(D\&E)), demonstrating the effectiveness of the  metal trace replacement component visually.

\begin{figure*}[t]
	\centering
	\includegraphics[width=1.0\linewidth]{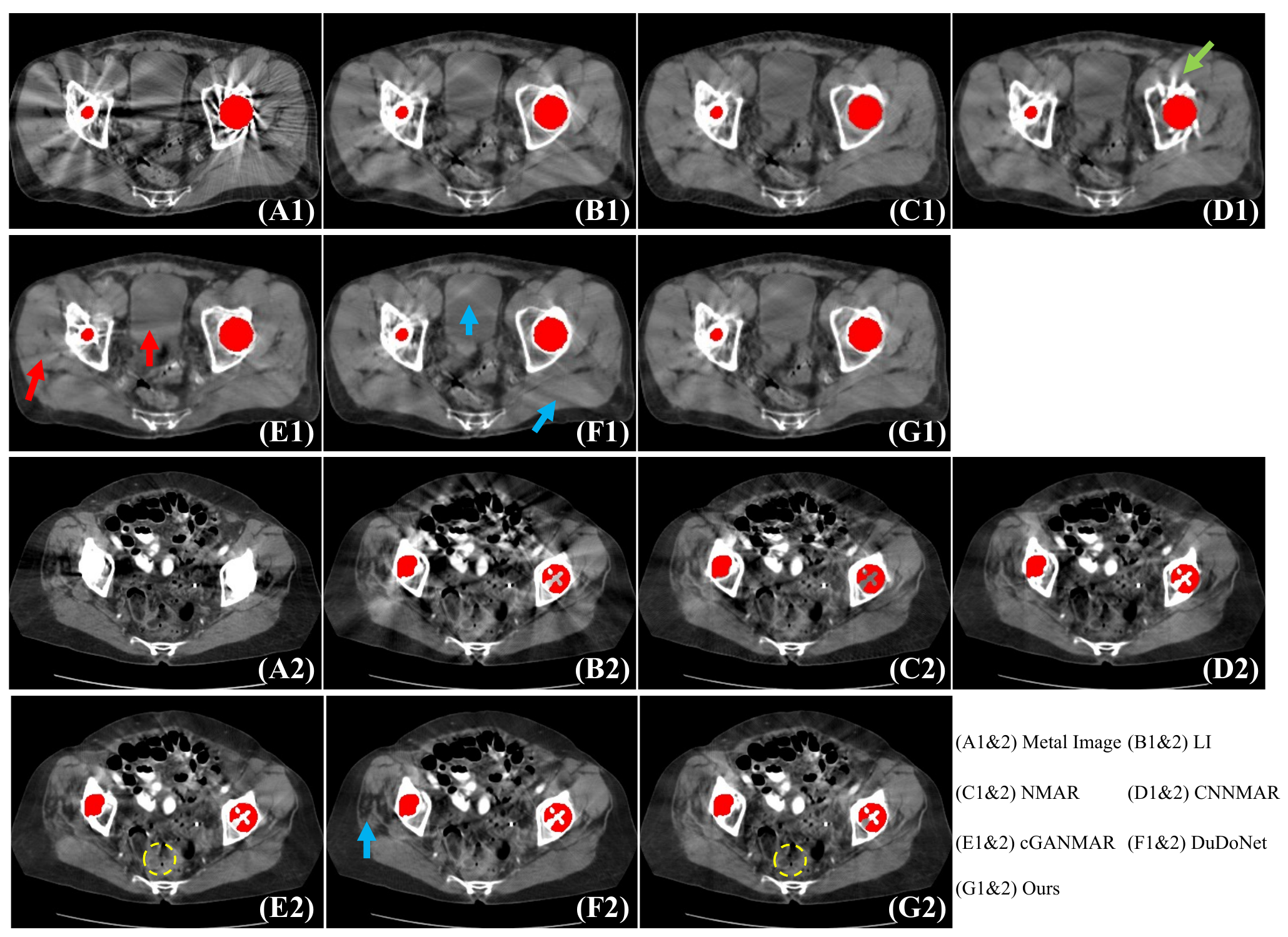}
	\caption{Visual results of different methods on clinical data with real metal artifacts. The figures are shown in [-160 310] HU.}
	\label{fig:realvisual}
	\centering
\end{figure*}

\subsection{Comparison with Other Methods}
We  compare our method with baseline methods LI~\cite{kalender1987reduction} and NMAR~\cite{meyer2010normalized}, which are widely used in MAR.
We also compare our method with the recently developed deep-learning methods CNNMAR~\cite{zhang2018convolutional} and cGANMAR~\cite{wang2018conditional}, and the state-of-the-art method DuDoNet~\cite{lin2019dudonet}.
We re-implement DuDoNet~\cite{lin2019dudonet} and cGANMAR~\cite{wang2018conditional} for comparison. 

As shown in Table~\ref{table:deeplesion}, the deep-learning-based methods CNNMAR and cGANMAR outperform the conventional MAR methods on both RMSE and SSIM metrics, showing the advantages of deep neural networks.
When compared with CNNMAR and cGANMAR, the DuDoNet achieves low RMSE and comparable SSIM values as it integrates both sinogram completion and image refinement to reduce  metal artifacts.
Our method further reduces RMSE with 3.80 HU and achieves slightly better SSIM on the DeepLesion data compared with DuDoNet.
The reason may be that the final output image of our method is directly reconstructed from the completed sinogram, so that it can better preserve the original intensity values. 

Fig.~\ref{fig:visual1} shows the visual comparisons of all the involved methods on two DeepLesion sample images from two organs, \ie, chest and abdomen \revise{(Readers can find much larger figures in the supplementary materials)}.
From (A1 - H1) of the chest and (A2 - H2) of the abdomen, we can observe that severe streaking artifacts are visible in the FBP-reconstructed results (B1, B2).
After correction, all the involved methods can alleviate the streaking artifacts to some extent; however, they differ in the degree of restoration of image details.
In conventional interpolation-based methods LI and NMAR (C1, D1), amounts of low-density shadows are mixed in the soft tissues, \eg, fat.
cGANMAR (F1) introduces some new high-density shadows around the spine marked by the green arrow.
CNNMAR (E1) and DuDoNet (G1) change the gray distribution of the main pulmonary artery marked by the red arrow.
Also, some blood capillaries can be preserved well by cGANMAR, DuDoNet and our methods.
Our method (H1) is most closet to the reference CT image (A1) compared with other correction methods and the same trend can be observed in the experimental result on the abdomen.
Here, macroscopic streaking artifacts along the edge of the metal are marked by the blue dotted ellipse in (C2, F2, G2) and the artifacts around the bone structure are marked by the blue arrow in (D2, E2).

\begin{figure*}[t]
	\centering
	\includegraphics[width=1.0\linewidth]{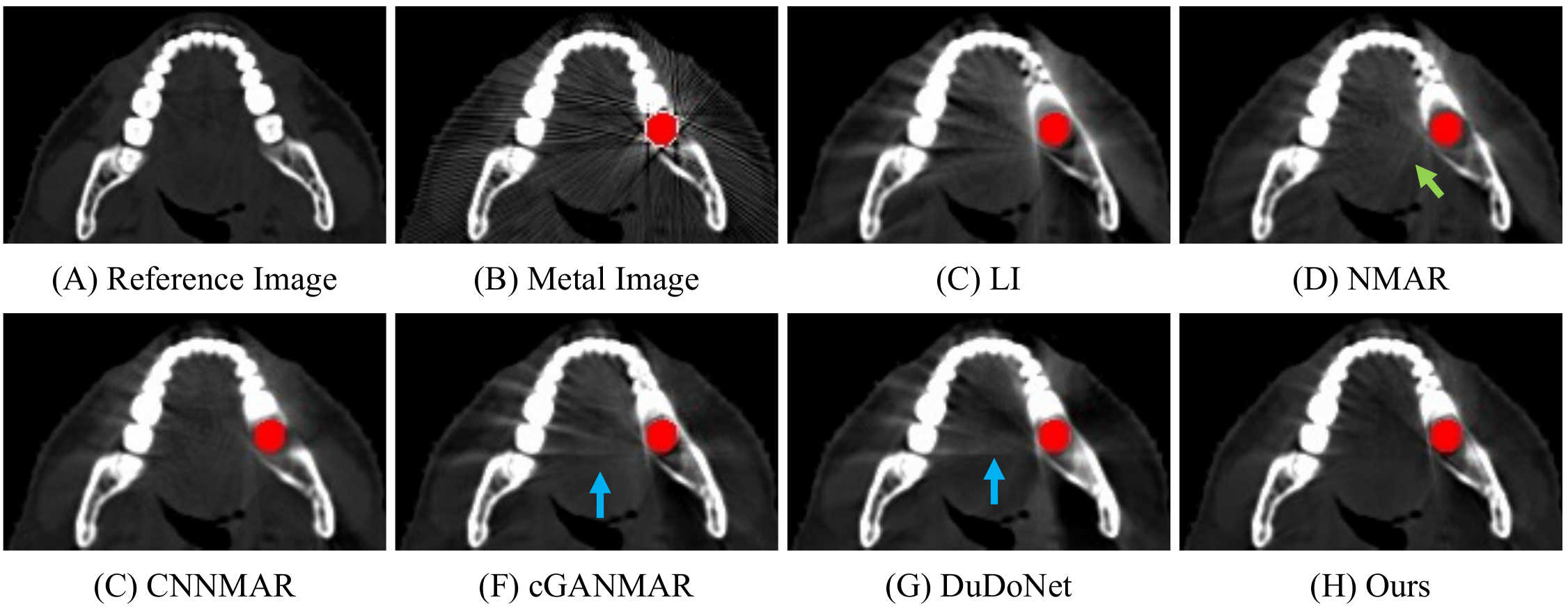}
	\caption{The visual comparison of different methods on head CT images. The simulated dental fillings are remarked as red. The figures are shown in [-200 1400] HU.}
	\label{fig:visualhead}
	\centering
\end{figure*}

\subsection{Experiments on CT Images with Real Artifacts}
The original metal-corrupted sinogram are difficult to access in real clinical scenario, we follow the procedure in~\cite{lin2019dudonet} to evaluate our framework on real artifact data.
In practical, we collected some clinical metal-affected CT images from the DeepLesion dataset and a local hospital and employed a  thresholding-based segmentation method  (\ie, 2000 HU) to segment the metal mask from the clinical CT images.
Then we conducted forward projection on the real metal-affected CT images and metal masks to acquire the metal-corrupted sinogram $S_{ma}$ and the metal trace region $Tr$, respectively.
Finally, we generate $S_{LI}$  from $S_{ma}$ and $Tr$ and fed it into our framework to get the final results.
Fig.~\ref{fig:realvisual} presents the visual results on two clinical CT images with bilateral hip prostheses. 
A severe dark strip between two hip prostheses can be observed in the original metal images; see Fig.~\ref{fig:realvisual}(A1\&A2). 
Compared with the original metal images, our method can further alleviate metal artifacts for both two samples; see Fig.~\ref{fig:realvisual}(G1\&G2).
For the first sample, we can see that the CNNMAR cannot effectively remove the strong artifacts near the metals as indicated by the green arrow in Fig.~\ref{fig:realvisual}(D1).
Our method reduces more artifacts than cGANMAR (see the red arrows in Fig.~\ref{fig:realvisual}(E1)) and DuDoNet (see the blue arrows in Fig.~\ref{fig:realvisual}(F1)). 
For the second sample, the CGANMAR slightly changes the anatomical structure while our method can preserve it; see the yellow circles in Fig.~\ref{fig:realvisual}(E2\&G2). 
Also, our method can better correct the dark strip artifacts than DuDoNet; see the blue arrow in Fig.~\ref{fig:realvisual}(F2). 
Since the second metal image sample has already been corrected by some MAR algorithms, the metal artifacts are a little mild and the differences among Fig.~\ref{fig:realvisual}(E2, F2, \&G2) are not obvious than the first sample. 

\subsection{Experiments on Different Site Data}
The training images from DeepLesion are mainly from the abdomen and thorax.
We also collected other head CT images from MICCAI 2015 Head and Neck Auto Segmentation Challenge dataset~\cite{raudaschl2017evaluation} and evaluated our well-trained model on these head CT images to show the  generalization capability of our method over different organ data.
The original size of the new head CT image is $512\times512$ and we adopted the same procedure to simulate the metal artifacts.
We then directly employ the network trained with DeepLesion data to process those head CT images.
We compare our method with other methods in Fig.~\ref{fig:visualhead}.
From the results, we can see that the original metal images suffer from severe artifacts due to the metal implants.
The conventional interpolation-based methods LI and NMAR can reduce the artifacts to some extend while they still suffer from severe artifacts, especially in the vicinity of the tooth.
Although trained with images from other sites, our method demonstrates an outstanding capacity to reduce artifacts on new head CT images and generates better results than other deep-learning-based methods, indicating that the proposed method has the potential to process different site data.

\subsection{Comparison with More Recent Methods}
We also compare our method with \revise{the} more recently proposed method ADN~\cite{liao2019adn}, DuDoNet++~\cite{lyu2020dudonet}, and our previous Deep sinogram completion work~\cite{yu2020deep}. 
We directly evaluated the ADN method on the simulated DeepLesion data with the public model provided by the authors. 
For the DuDoNet++, since the authors did not provide the public code and trained model, we reimplemented it and trained the network with our simulated training data.  
Table~\ref{Table:recent} shows the comparison results. It is observed that the generative adversarial network (GAN)-based method ADN produces relatively good performance on the SSIM metric, while the RMSE metric is much lower than our method and other methods. The reason is that the adversarial loss focuses more on the distribution similarity (\ie, appearance) between the generated metal-reduced image and other metal-free images, while there are no hard constraints for the intensity discrepancy in the objective function. 
The performance of DuDoNet++ is better than ADN, while it is still inferior to our method. 
Note that our previous deep sinogram completion work is based on supervised learning and requires anatomically identical metal-corrupted and metal-free data to train the network. Our new work adopts the self-supervised learning strategy and utilizes less supervised signals in network training so that the absolute performance of our method is lower than the previous deep sinogram completion work. However, these two results cannot be directly compared. 
\section{DISCUSSION}
\label{sec:discussion}
Metal artifact reduction  (MAR) is a long-standing yet challenging problem in CT imaging.  Although many methods have been developed, there are still no standard solutions in clinical practice~\cite{gjesteby2016metal}.
With the development of deep learning techniques in medical image analysis problems, the data-driven technique provides a promising direction to tackle this challenging problem.
Many recent works propose to use convolutional neural networks to address \revise{the} MAR problem.
However, most of these methods are supervised methods (\eg, CNNMAR~\cite{zhang2018convolutional} and our previous work~\cite{yu2020deep}), which require synthesized metal-free and metal-inserted CT training pairs.
Since the synthesized metal artifacts may not accurately model the real clinical artifacts, the performance of networks trained with synthesized training pairs would degrade in real clinical applications~\cite{liao2019adn}.
Therefore, in this work, we aim to develop self-supervised deep learning techniques for MAR, without relying on the synthesized training pairs.
Our framework is trained with metal-free sinograms and metal traces, and thus alleviates the need for metal-free and metal-inserted CT image training pairs, which is more suitable for real clinical practice.

The proposed self-supervised cross-domain learning framework solves MAR in both sinogram and image domains, which shares a similar idea with DuDoNet~\cite{lin2019dudonet}. However, our framework has some differences.
In DuDoNet, the authors directly adopt the CNN-refined images as final MAR images.
Although the CNN-refined images significantly reduce the metal artifacts, it would generate some failure cases, and it is difficult to remove those mild artifacts (see Fig.~\ref{fig:visual1}(G2) and Fig.~\ref{fig:realvisual}(F1) in the revised manuscript).
Therefore, inspired by previous prior-image-based MAR methods, we incorporate the reconstruction model (\ie, metal trace replacement) as the last step to preserve fine details and avoid resolution loss, often ignored in other deep-learning-based methods, like DuDoNet (see comparisons in Fig.~\ref{fig:visual1}).
Furthermore, instead of directing training a sinogram inpainting network, we design a novel FBP \revise{reconstruction} loss to encourage the network to generate \revise{more perfect sinogram} completion results by minimizing the difference of a pair of completion results from the sinogram. 
This strategy is different from directly minimizing the difference between the FBP constructions with image ground truth. The peered FBP construction provides ``soft supervision”, not the ``hard supervision” from the \revise{reference image}~\cite{hinton2015distilling} to encourage geometry-consistent sinogram completion so that it can ease the network learning. 
Moreover, with our proposed FBP \revise{reconstruction} loss, the two branches could learn in a mutual manner and benefit each other~\cite{zhang2018deep}.
This strategy can provide strong regularization for network learning, and the mechanism under it is very similar to the positive pair construction in the contrastive loss~\cite{chen2020simple}, 
which achieves great success in the self-supervised learning field.
Based on these new designs, our method outperforms DuDoNet with 3.80 HU, as shown in Table~\ref{table:deeplesion}.
At the current stage of this study, the network learning and metal trace replacement are conducted separately.  
In the future, we will investigate how to formulate the metal trace replacement as a network component, so that we can  train the  joint network and metal trace replacement component in an end-to-end manner and the two procedures can benefit from each other.

\begin{table} [t]
	\centering
	\caption{Comparison with more recent methods on DeepLesion simulation dataset. * denotes supervised learning methods.}
	\label{Table:recent}
	
	\begin{tabular}{p{4.7cm}<{\centering}|p{1.9cm}<{\centering}|p{1.9cm}<{\centering}}
		\toprule[1pt]
		Method & RMSE (HU) & SSIM\\ \hline	
		ADN~\cite{liao2019adn}				     	 &89.34     &0.9764     		\\	
		DuDoNet++*~\cite{lyu2020dudonet}   	&38.29     &0.9770    	\\
		Deep sinogram completion*~\cite{yu2020deep}	 &31.15     &0.9784  \\
		\textbf{(Ours)}            &\textbf{34.20}     &\textbf{0.9773}    \\
		\bottomrule[1pt]
	\end{tabular}
\end{table}

Our self-supervised learning framework needs to take the metal-free sinogram and metal trace as input to train the model. Since it is difficult to acquire the real sinogram data from CT equipment, we simulate sinogram data from clinical CT images in DeepLesion dataset to train the network. 
Suppose we can access to the sinogram data taken from a CT machine, then we can train our network with real sinogram data when deploying our framework into clinical scenarios. 
Considering that simulating massive physical-constrained metal trace (or metal artifacts) is very difficult and not the main goal of this work, in our current work, we generate the metal trace region from randomly collected metal masks, which would lead to some invalid metal implants simulation (\eg, metal in lungs).
Since our network is trained with massive simulated data (the different combinations of sinogram and metal trace), those invalid simulations could act as “regularization” to the network training and would not largely influence the performance of testing samples. 
We regard how to simulate more physical-constrained metal trace regions as our future work. 

In our experiments, the simulated CT images and metal masks in training and validation sets are different, showing that our method would handle new metal masks and new CT slices.
Although our model is trained with abdomen and thorax CT samples, as shown in the experiments, our model has the potential to process other site data, \eg, head CT images.
In our current study, we use the same simulated scanner geometry for training and validation. 
In the future, we can test our trained model with different scanner geometry to see the robustness of our method. 
Also, we may train the network with more diverse data from different sites so that the trained model can better process different metal artifact types.

One limitation of our current work is that we only evaluate our method on simulated sinogram data not raw projections in Radon space. In the future, we will collaborate with our clinical collaborators to collect raw projections in hospitals and evaluate the performance of our method for clinical applications.
One important issue when deploying our framework into clinical practice is to obtain an accurate \revise{metal mask}.
In our work, we following previous works and adopt a thresholding method to segment metal masks from reconstructed metal-affected CT images.
As deep learning has achieved good performance on various medical image segmentation tasks~\cite{ronneberger2015u,litjens2017survey}, we can incorporate these powerful deep-learning-based segmentation methods into our framework to accurately segment the metal masks.
Also, it is very promising to investigate how to conduct metal mask identification and metal artifact reduction simultaneously, e.g., combining the binary reconstruction~\cite{wang2010binary,meng2010sinogram} into our framework.
Another limitation of our current work is that we only consider 2D slices in our method. Extending our framework to 3D scans is an interesting and important topic and we regard it as our future work.


\section{Conclusion}
\label{sec:conclusion}
We present a generalizable deep learning method for metal artifact reduction in CT images, which combines the merits of sinogram and image domain learning.
To alleviate the need for anatomically identical metal-free and metal-affected CT image pairs, we formulate our method as a self-supervised learning framework and design a novel FBP consistency loss to encourage the network to generate geometry-consistent completion results.
The final MAR image is directly acquired from the sinogram with the FBP reconstruction, preserving the fine details of the original structure.
Our proposed self-supervised learning framework would advance the deep learning technique for high-quality CT imaging. 
%

%
\IEEEpeerreviewmaketitle

\ifCLASSOPTIONcaptionsoff
  \newpage
\fi



\bibliographystyle{IEEEtran}
\bibliography{refs}
\end{document}